\documentclass[twocolumn]{emulateapj}
\usepackage{color}
\usepackage{amsmath}

\newcommand{\erfc}{{\rm{}erfc}}

\newcommand{\rr}{\mathbf{r}}
\newcommand{\hh}{\mathbf{h}}
\newcommand{\vv}{\mathbf{v}}

\newcommand*{\defeq}{\stackrel{\text{def}}{=}}

\newcommand{\ctxt}{\mathcal{C}}  

\newcommand{\eind}{\epsilon}  
\newcommand{\edata}{D_\eind}  
\newcommand{\like}{\mathcal{L}}  
\newcommand{\flike}{\ell}  
\newcommand{\dlike}{m}  
\newcommand{\mlike}{\mathcal{M}}  

\newcommand{\flux}{f}
\newcommand{\fest}{\hat{\flux}}  
\newcommand{\fsig}{\sigma}  
\newcommand{\drxn}{\mathbf{n}}
\newcommand{\dest}{\hat{\mathbf{n}}}  
\newcommand{\dsig}{\delta}  

\newcommand{\dtxn}{\mathcal{D}}  
\newcommand{\ndtxn}{\mathcal{N}}  
\newcommand{\fth}{\flux_{\rm th}}

\newcommand{\npd}{\nu}  

\newcommand\enote[1]{ {\color{red}%
\marginpar[\raggedleft\large $\blacktriangleright$]%
{\raggedright\large $\blacktriangleleft$} %
{\large $\langle\langle\langle$}{\sl #1}{\large  $\rangle\rangle\rangle$} }}

\newcommand\tamas[1]{{\color{blue}#1 [Tamas]}}

\begin{document}

\title{Faint Object Detection in Multi-Epoch Observations\\ via Catalog Data Fusion}
\journalinfo{Submitted to Apj}

\author{ Tam\'{a}s Budav\'{a}ri, Alexander S.\ Szalay}
\affil{Dept.\ of Physics and Astronomy, The Johns Hopkins University, 3400 North Charles Street, Baltimore, MD 21218}
\and
\author{Thomas J. Loredo}
\affil{Cornell Center for Astrophysics and Planetary Science, Cornell University, Ithaca, NY 14853}

\shortauthors{}
\shorttitle{}

\begin{abstract} 
Observational astronomy in the time-domain era faces several new challenges. 
One of them is the efficient use of observations obtained at multiple epochs. 
The work presented here addresses faint object detection with multi-epoch data, and describes an incremental strategy for separating real objects from artifacts in ongoing surveys, in situations where the single-epoch data are summaries of the full image data, such as single-epoch catalogs of flux and direction estimates (with uncertainties) for candidate sources.
The basic idea is to produce low-threshold single-epoch catalogs, and use a probabilistic approach to accumulate catalog information across epochs; this is in contrast to more conventional strategies based on co-added or stacked image data across all epochs.
We adopt a Bayesian approach, addressing object detection by calculating the marginal likelihoods for hypotheses asserting there is no object, or one object, in a small image patch containing at most one cataloged source at each epoch.
The object-present hypothesis interprets the sources in a patch at different epochs as arising from a genuine object; the no-object (noise) hypothesis interprets candidate sources as spurious, arising from noise peaks.
We study the detection probability for constant-flux objects in a simplified Gaussian noise setting, comparing results based on single exposures and stacked exposures to results based on a series of single-epoch catalog summaries.
Computing the detection probability based on catalog data amounts to generalized cross-matching: it is the product of a factor accounting for matching of the estimated fluxes of candidate sources, and a factor accounting for matching of their estimated directions (i.e., directional cross-matching across catalogs).
We find that probabilistic fusion of multi-epoch catalog information can detect sources with only modest sacrifice in sensitivity and selectivity compared to stacking.
The probabilistic cross-matching framework underlying our approach plays an important role in maintaining detection sensitivity, and points toward generalizations that could accomodate variability and complex object structure.
\end{abstract}

\keywords{methods: statistical --- surveys --- catalogs --- photometry --- astrometry}

\section{Introduction}
\label{sec:intro}
\noindent
In the era of time-domain survey astronomy, dedicated telescopes scan the sky every night and strategically revisit the same area several times.
The raw data are images, but surveys commonly provide, not only image data, but also \emph{catalogs}, summaries of the image data that aim to enable a wide variety of studies without requiring users to analyze raw or processed image data.
Catalogs typically report object properties, based on algorithms that detect sources in images with a measure of statistical significance above some threshold, chosen so that the resulting catalog is likely to be highly pure (i.e., with few or no spurious sources).
 
The question we address in this paper is how to combine information from a sequence of independent observations to maximize the ability to detect faint objects at or near a chosen detection threshold, ameliorating the data explosion due to false detections from a lower threshold that would be required by a suboptimal method.
Focusing on the faint objects that typically dominate the collected data is an important and timely problem for a number of ongoing surveys and vital for planning the next-generation data processing pipelines.

There are two different ways one can approach the problem. 
A traditional, resource-intensive approach is to wait until all observations are completed, performing detection by stacking the multi-epoch image data (with potential complications related to registration, resampling, and point spread function matching).
An optimal procedure for threshold-based detection with image stacks was introduced by \citet{chisq}.
Once a master object catalog is produced from the stacked images, time series of source measurements are created by forced photometry at the master catalog locations.

An alternative (non-exclusive) approach is to perform source detection independently for each observation, producing a catalog of candidate sources at each epoch \citep{BS14-ADASS}.
The detection threshold may be different for each epoch. 
Interim object catalogs may be produced by analysis of the series of overlapping source detections potentially associated with a single object using any available data; a final catalog would be built using the catalogs from all epochs.
Of course, a catalog based on data from many epochs should be able to include many dim sources that escape confident detection in single-epoch or few-epoch catalogs.
To enable construction of a deep multi-epoch catalog, the single-epoch catalogs must report information for candidate sources with relatively low statistical significance; i.e., the single-epoch catalogs must have reduced purity.
If we set the single-epoch threshold too high, there will be too few detections; we will not have well-sampled time series for dim sources, and the final catalog will be too small.
If, on the other hand, we set the threshold too low, the single-epoch catalogs will be overwhelmed with (mostly false) detections that were seen only once, requiring wasteful expenditure of storage and computing resources for constructing multi-epoch catalogs.
An optimal threshold might preserve the size and quality of the final catalog, while enabling users to build interim catalogs on-the-fly, potentially tailored to specific, evolving needs.

We here address the second alternative, considering how best to accumulate evidence from possibly marginal detections while the observations are in progress, to prune spurious source detections but keep the sources associated with genuine objects.
The study presented here is exploratory, to establish the basic ideas and provide initial metrics for studying the feasibility of the incremental approach.
To make the analysis analytically tractable and the results easy to interpret, we restrict ourselves to an idealized setting; we will present a more general and formal treatment in a subsequent paper.

\section{Detection Probabilities} 
\label{sec:det}

\noindent

We adopt the terminology of LSST and other time-domain synoptic surveys, using \emph{source} to refer to single-epoch detection and measurement results, and \emph{object} to refer to a unique underlying physical system (e.g., a star or galaxy) that may be associated with one or more sources.
(Here we limit ourselves to objects that would appear as a single source.)
For simplicity, we consider observations in a single band unless stated otherwise.

Consider an object with constant flux $f$ and direction $\drxn$ (a unit-norm vector on the sky). 
At each epoch $\eind$, analysis of the image data $\edata$ corresponding to a small patch of sky of solid angle $\Omega$ produces a \emph{source likelihood function} (SLF) for the basic observables, flux $\flux$ and direction $\drxn$, of a candidate source in the patch.
The SLF is the probability for the data as a function of the (uncertain) values of the observables,
\begin{equation}
\like_\eind(\flux,\drxn) \equiv p(\edata|\flux,\drxn,\ctxt),
\label{like-def}
\end{equation}
where $\ctxt$ denotes various contextual assumptions influencing the analysis,
e.g., specification of the photometric model and information about instrumental and sky backgrounds.
(Since $\ctxt$ is common to all subsequent probabilities, we henceforth consider it as implicitly present.)
For example, if photometry is done via point spread function (PSF) fitting with weighted least squares, and if the noise level and backgrounds are known, then it may be a good approximation to take $\like_\eind(\flux,\drxn) \propto \exp[-\chi^2(\flux, \drxn)/2]$, where $\chi^2(\flux, \drxn)$ is the familiar sum of squared weighted residuals as a function of the source flux and direction.

We consider a catalog at a given epoch to report summaries of the likelihood functions for candidate sources that have met some detection criteria.
The most commonly reported summaries are best-fit fluxes (or magnitudes) with a quantification of flux uncertainty (typically a standard error or the half-width of a 68\% confidence region), and, separately, best-fit sky coordinates with an uncertainty for each coordinate.%
We here take these summaries to correspond to a factored approximation of the source likelihood function,
\begin{align} \label{eq:eplike}
\like_\eind(\flux, \drxn)
  &\equiv p(\edata|\flux, \drxn, H_1)\nonumber\\
  &= \flike_\eind(\flux)\, \dlike_\eind(\drxn),
\end{align}
where the epoch-specific flux factor, $\flike_\eind(\flux)$, is a Gaussian with mode $\fest_\eind$ (the catalog flux estimate at epoch $\eind$) and standard deviation $\fsig_\eind$, and the direction factor, $\dlike_\eind(\drxn)$, is an azimuthally symmetric bivariate Gaussian with mode $\dest_\eind$, and standard deviation $\dsig_\eind$.%
\footnote{Care should be taken in reporting and interpreting $\dsig_\eind$, because direction is a two-dimensional quantity.
If $\dsig_\eind$ is the single-coordinate standard deviation, the angular radius of a 68.3\% (``$1\sigma$'') confidence region or flat-prior credible region is $\approx 1.52\dsig_\eind$.}
This may be a rather crude approximation; we will address it further elsewhere, here merely noting that it is implicitly adopted for most survey catalogs.
For simplicity, we take the flux factors to have the same standard deviation at all epochs,  $\fsig_\eind = \sigma$.



We adopt a simple source detection criterion: a candidate source with a flux likelihood mode $\fest$ larger than a single-epoch threshold value, $\fth$, is deemed a detection.
The probability for detection in a single-epoch catalog is the probability that source with true flux $f$ will produce a single-epoch measurement $\fest_\eind$ that falls above the threshold.
This probability is just the integral of the Gaussian flux likelihood function above the threshold, which we denote by
\begin{equation}\label{eq:pf}
P_f \defeq P(>\!\fth|f) = \frac{1}{2}\,\erfc\left(\frac{\fth\!-\!f}{\sigma\sqrt{2}}\right),
\end{equation}
where $\erfc(\cdot)$ is the complementary error function.
 
\begin{figure}
\epsscale{1.2}
\plotone{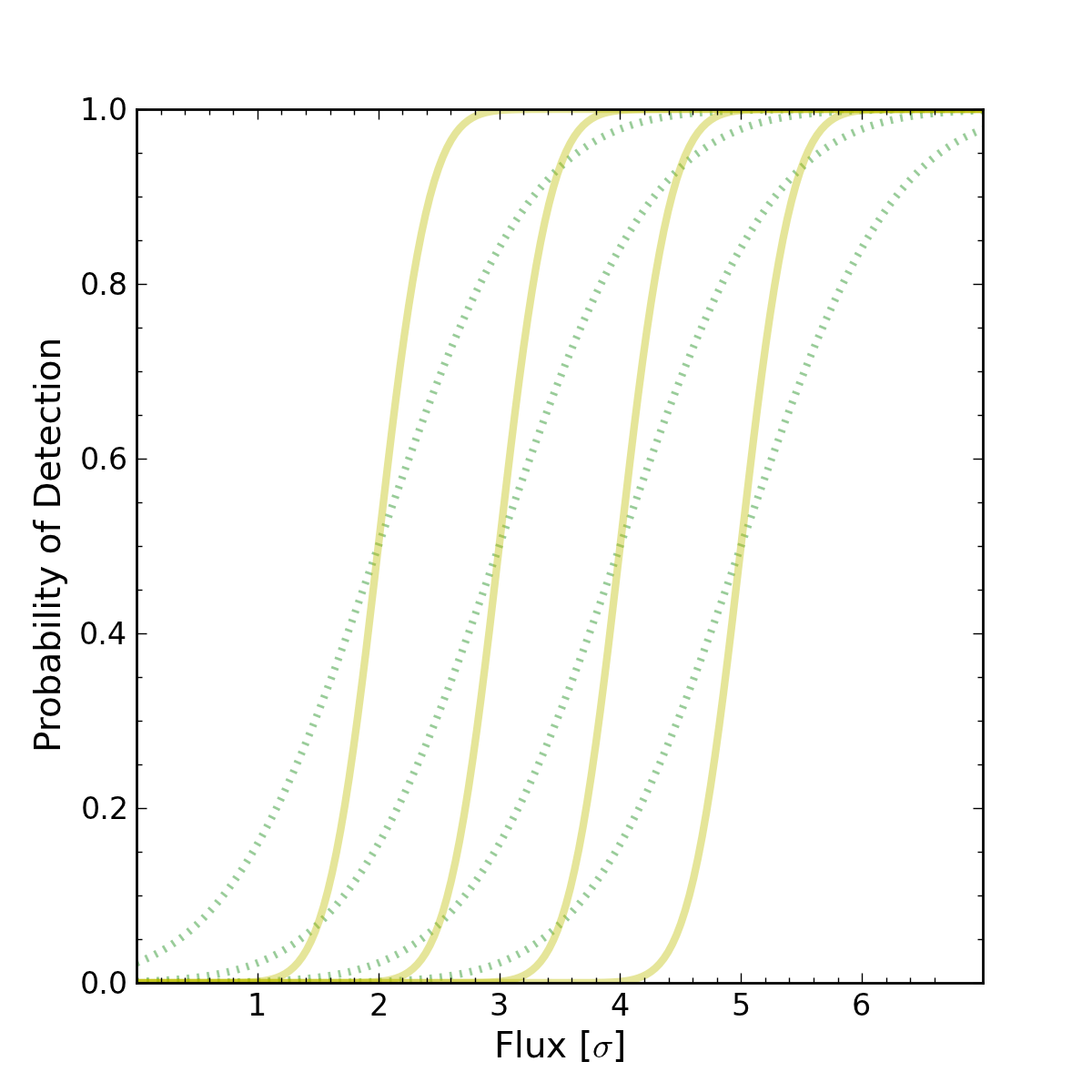}
\caption{The detection probability is shown in different scenarios as a function of the true flux in $\sigma$ units. The green dotted lines illustrate single-exposure cases with different thresholds that take values of 2, 3, 4, and 5$\sigma$ ({\it{}from left to right}). The yellow solid lines are the same for stacks with \mbox{$k\!=\!9$} observations, at the same flux thresholds.}
\label{fig:1}
\end{figure}

For comparison, consider detection probabilities in the case of stacked exposures from $k$ observations.
We assume that the objects are stationary and have a constant flux, and that the dominant source of noise is still the sky, so the relative noise is reduced by $\sqrt{k}$ after stacking.
For a stacked exposure flux threshold $f_S$, the probability for detection is
\begin{equation}
P'_f \defeq \frac{1}{2}\,\erfc\left(\frac{f_S\!-\!f}{\sigma\sqrt{2/k}}\right).
\end{equation}
Figure~\ref{fig:1} displays the detection probability as a function of true object flux for single-epoch and stacked data, for various choices of the single-epoch and stacked thresholds.
The dotted green lines represent the single-exposure situation $P_f$ as a function of the true flux in $\sigma$ units for detection thresholds of 2, 3, 4, and 5$\sigma$.
Similarly the solid yellow lines correspond to the stacked detections with \mbox{$k\!=\!9$} exposures.

Consider two curves corresponding to the same threshold, so $f_S = \fth$.
The probability for detection at $f=\fth$ is 50\% for both a single exposure and stacked exposures.
But the curve for stacked exposures is much steeper, with a higher probability for detecting sources brighter than $\fth$, and a lower probability for detecting sources dimmer than $\fth$.
That is, when constructed with a common threshold, the catalog built from stacked data will be more complete above threshold, and will more effectively exclude sources with true flux below threshold.

\subsection{Multiple detections and non-detections}
\begin{figure}
\epsscale{1.2}
\plotone{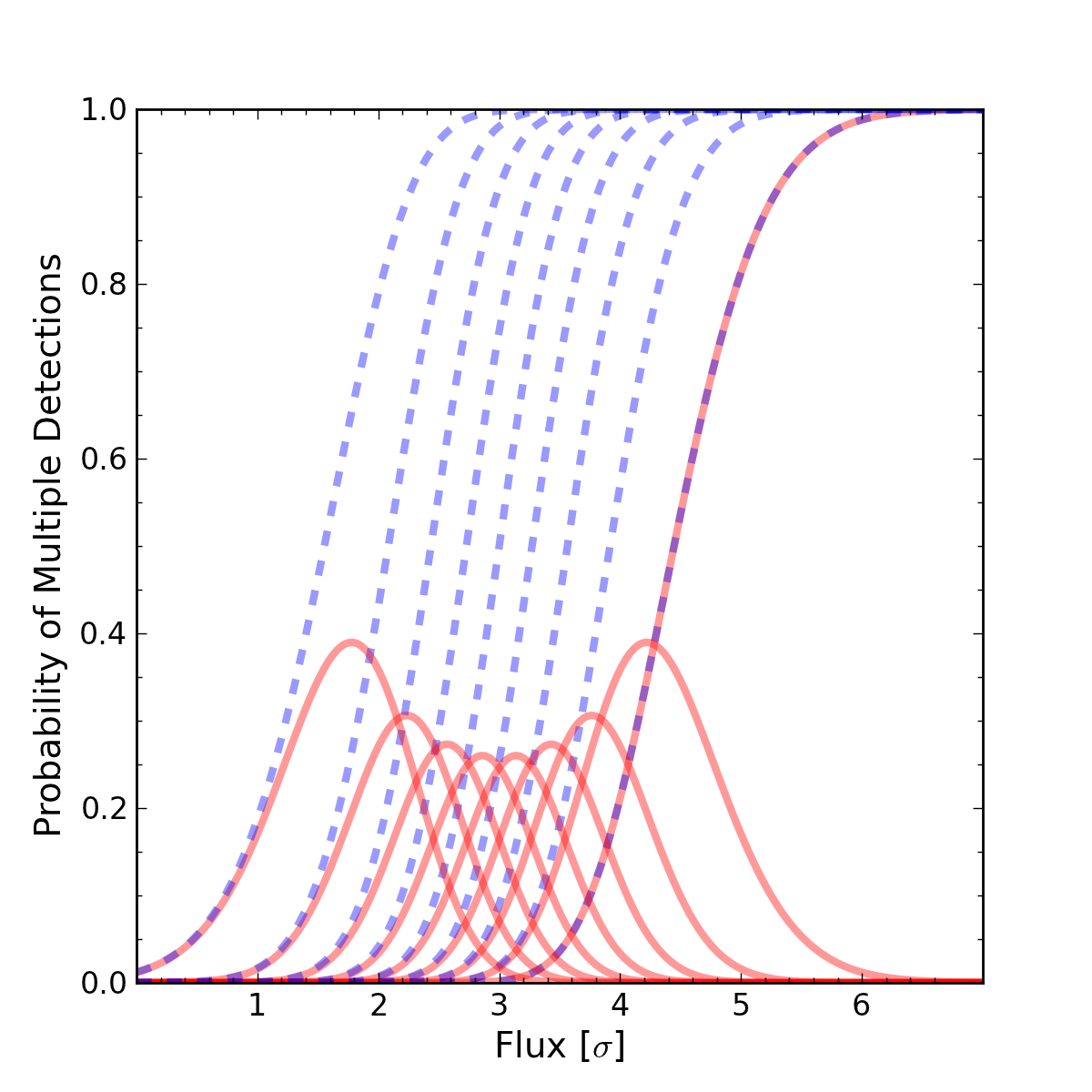}
\caption{The detection probability in multiple observations is shown here as function of the true flux. Assuming \mbox{$k\!=\!9$} total available observations, the {\it{}solid red curves} show the probability of the object appearing in exactly 1, 2, 3, etc., up to 9 observations ({\it{}from left to right}). Similarly, the {\it{}dashed blue lines} correspond to the 1+, i.e., 1 or more, 2+, 3+, etc., cases.}
\label{fig:2}
\end{figure}

\noindent
Faint sources will not always be detected.
The probability for making $n$ detections among \mbox{$k=n+m$} observations follows the binomial distribution, giving the multi-epoch detection probability,
\begin{equation} \label{eq:binomial}
 P(n|k,f) = {k \choose n}\ P_f^n\,\big(1\!-\!P_f\big)^{k-n}.
\end{equation}
An interesting quantity is the probability that an object would lead to source detections in $n_0$ or more observations.
This is simply the sum
\begin{equation}
P(n\!\geq{}\!n_0|k,f) = \sum_{n=n_0}^k {k \choose n}\ P_f^n\,\big(1\!-\!P_f\big)^{k-n}
\end{equation}
(this can be expressed in terms of the incomplete beta function).
In Figure~\ref{fig:2} we plot these probabilities as a function of $f$ (again in $\sigma$ units) for \mbox{$k$=9} observations.
From left to right, the solid red curves show the probability for detecting an object of given flux in exactly 1, 2, etc., up to 9 observations.
Similarly the dashed blue curves correspond to cases $1+$ (1 or more), $2+$, and so on.
(Note that the functions coincide for the case $n=n_0=k=9$.)

\begin{figure}[t]
\epsscale{1.2}
\plotone{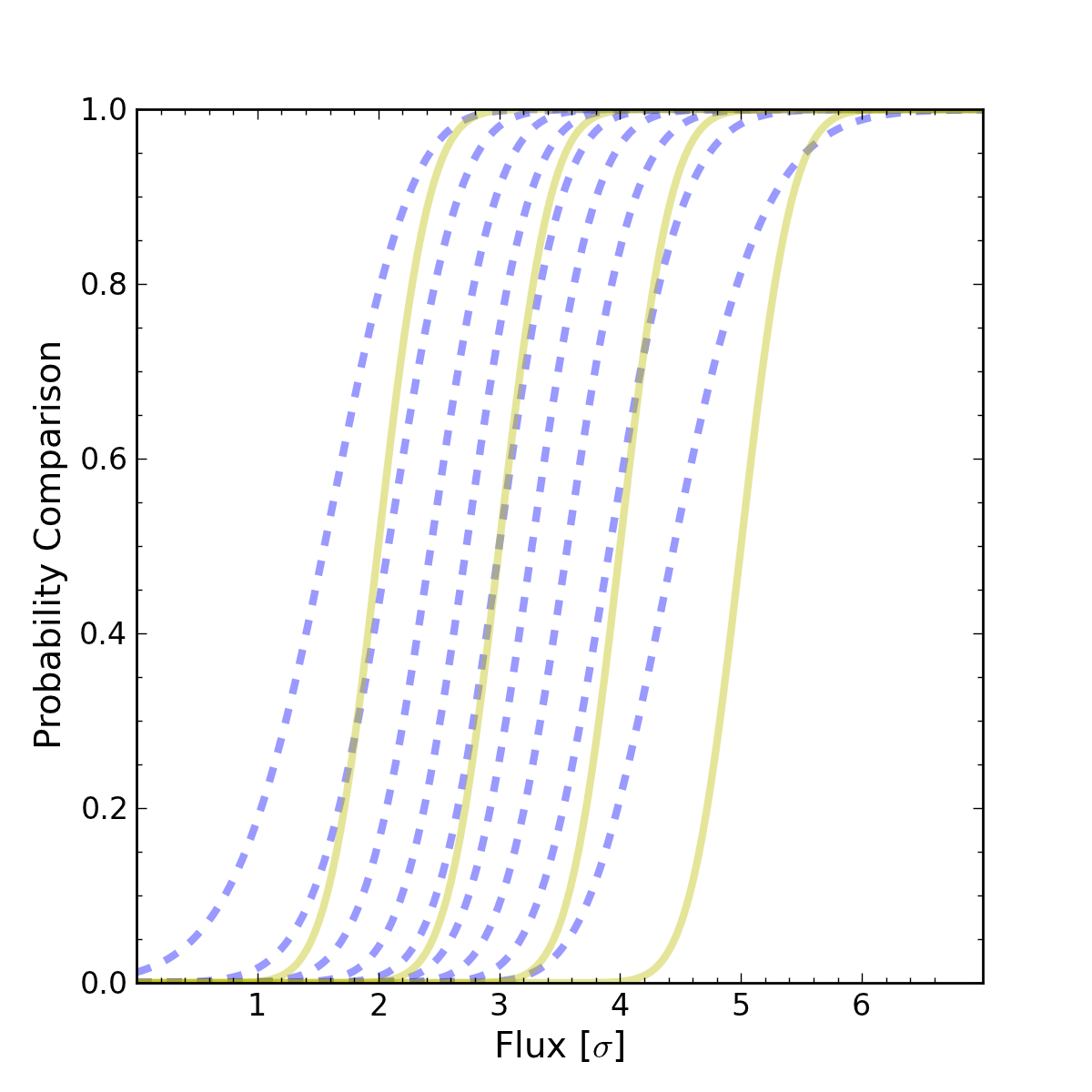}
\caption{The comparison of the probabilities plotted previously in Figures~\ref{fig:1} and \ref{fig:2} reveals the similarities of the alternative methods. The curves of the stacked cases ({\it{}solid yellow lines}) and the summed up binomials ({\it{}dashed blue lines}) follow similar trends in the usual regime of parameter space. In particular, we highlight the remarkable agreement of the 3$\sigma$ curve for the stack detections ({\it{}solid yellow line second from the left}) and the 5+ sum of the binomials ({\it{}dashed blue in the middle}).}
\label{fig:3}
\end{figure}

Figure~\ref{fig:3} compares detection probability curves for the stacked exposure case (solid yellow curves, as in Figure~\ref{fig:1}) and the multi-epoch, $n>n_0$ detection case (dashed blue curves, as in Figure~\ref{fig:2}).
For a particular stacked exposure case, we see there is a multi-epoch case whose detection probability curve displays very similar performance.
For example, the $3\sigma$ stacked exposure curve is very similar to the multi-epoch $5+$ detection case.
This indicates that collecting sources with $5+$ detections from \emph{single-epoch} $3\sigma$ catalogs is nearly equivalent in terms of catalog completeness and purity to producing a separate, new $3\sigma$ stacked exposure catalog.
%


Analyzing the single-epoch catalogs has a number of advantages.
It can be implemented in an incremental fashion that follows the schedule of the survey, and the time series data are readily available at a given time; there is no need to go back to old images and to performed forced photometry at locations that are revealed only in the final stack.

\section{Distinguishing Real Sources from Noise Peaks}
\noindent
The previous calculations addressed detectability of a source of known true flux, $f$.
In real-life scenarios, the problem is quite the opposite---we are presented with the observations and would like to understand the properties of the sources. 
In this context, our focus is on how one can reliably distinguish noise peaks from real sources. 
It is important to emphasize that we have more information than just the fact that a source has been detected; we also have flux measurements, at multiple epochs. 
Our approach is motivated by Bayesian hypothesis testing, where the strength of evidence for presence of a source is quantified by the posterior probability for the source-present hypothesis, or equivalently, by the posterior odds in favor of a source being present vs.\ being absent (the odds is the ratio of probabilities for the rival hypotheses).
The posterior odds is the product of prior odds and the data-dependent \emph{Bayes factor}.
The prior odds depends on population properties; it may be specified a priori when there is sufficient knowledge of the population under study, or learned adaptively by using hierarchical Bayesian methods (for examples of this in the related context of cross-identification, see \cite{B13-SCMAXMatch,L13-HierXMatch}).
Here we focus on the Bayes factor; we will address hierarchical modeling in a follow-up paper.

The Bayes factor is the ratio of marginal likelihoods for the competing hypotheses, one that claims that the sources are associated with a real object, and its complement that assumes there is just noise:
\begin{equation} \label{eq:Bfac}
B = \frac{\mlike_{\rm{}real}}{\mlike_{\rm{}noise}} .
\end{equation}
Each marginal likelihood, $\mlike_{\rm{}hyp}$, is the integral, with respect to all free parameters for the hypothesis, of the product of the likelihood function and the prior probability density for the parameters.

Let us now assume that out of $k$ observations, we measure $n$ detections with measured fluxes $\{\fest_\eind\}$.
We consider the two competing hypotheses separately.

\subsection{Real-object hypothesis}
\noindent
Let $\dtxn$ denote the set of indices for epochs with detections, and $\ndtxn$ denote the set of indices for epochs with nondetections:
\begin{equation}
\begin{split}
\dtxn  &\equiv \{\eind : \fest_\eind > \fth\},\\
\ndtxn &\equiv \{\eind : \fest_\eind \le \fth\}.
\end{split}
\label{D-ND-sets}
\end{equation}
For a candidate object with $n$ source detections among $k$ catalogs, the likelihood for a candidate true flux $f$ is
\begin{equation}
\like(f) = (1\!-\!P_f)^{k-n}\,\prod_{\eind \in \dtxn} \flike_\eind(\flux),
\end{equation}
where \mbox{$(1\!-\!P_f)$} is the probability of not detecting an object with true flux $f$, which happens \mbox{$(k\!-\!n)$} times, and $\flike_\eind(\cdot)$ is the flux likelihood function defined above (Gaussians with means equal to $\fest_\eind$).
The marginal likelihood for the real-object hypothesis is obtained by averaging over all possible true flux values, $f$.
For an object that is a member of a population with known flux probability density $\pi(f)$, the prior probability for $f$, used for the averaging in the marginal likelihood, is $\pi(f)$, so that
\begin{equation}
\mlike_{\rm{}real} = \int\!\!df\,\pi(f)\,L(f)
\end{equation}
which is a one-dimensional integral that can be analytically or numerically quickly evaluated.
(When the population distribution is not known a priori, it may be estimated via joint analysis of the catalog data for many candidate objects, within a hierarchical model, a significant complication that we will elaborate on elsewhere.)

\subsection{Noise peak hypothesis}\label{sec:peaks}
\noindent
The alternative hypothesis is that the detections are simply random noise peaks in the image.
The noise hypothesis marginal likelihood, $\mlike_{\rm noise}$, is the probability for the catalog data presuming no real object is present.%
\footnote{We are presuming the use of a fixed source detection algorithm, e.g., fixed apertures. 
If the algorithm is adaptive, its tuning parameters need to be accounted for in this probability.}

For epochs with a candidate source reported in the catalog, the datum is the flux measurement, $\fest_\eind$, and the relevant factor in the marginal likelihood is $\npd(\fest_\eind) \defeq p(\fest_\eind|\rm{Noise})$, the \emph{noise peak distribution}, evaluated at $\fest_\eind$.
This distribution will depend on the noise statistics for each catalog.

For epochs with no reported detection, we instead know only that $\fest_\eind \le \fth$, so the relevant factor is the fraction of \emph{missed} noise peaks,%
\footnote{The integral is over possible \emph{measured} flux values, $\fest_\eind$, not true flux; in the Gaussian regime assumed here the measured value may be negative, albeit with small probability.
The estimated flux would be constrained to be positive via the prior density, $\pi(f)$, which would multiply the flux likelihood when computing posterior flux estimates.}
\begin{equation} \label{eq:Q-noise}
Q_N \defeq \int_{-\infty}^{\fth}\!\!d\fest\,\npd(\fest).
\end{equation}
The probability for a false detection is then $P_N = 1 - Q_N$.

To compute these quantities comprising $\mlike_{\rm noise}$, we need to know the noise peak distribution, $\npd(\fest)$.
This distribution is not trivial to specify; it will depend both on the noise sources, and on the source detection algorithm.
Typically, a source finder performs a scan, identifying local peaks of the measured fluxes smoothed with a kernel, e.g., corresponding to a specified point source aperture.
Under the noise hypothesis, the source finder will be finding peaks of a smooth random field.
The locations and amplitudes of the peaks will form a point process, whose statistical properties can be analytically calculated \citep{adler,bbks,bond,kaiser}.
The most important consequence is that even though the underlying noise at the pixel level may be independent and Gaussian, the source finder output will correspond to sampling from a point process with a more complicated distribution of fluxes.
In particular, although the pixel-level noise distributions are symmetric (about the mean background), the distribution for (falsely) detected fluxes is skewed toward positive values.
The relevant calculation is presented in the Appendix.

Figure~\ref{fig:surface} shows the surface density of noise peaks as a function of the detection statistic $\fest$ (in $\sigma$ units), in the scenario when the sky noise is spatially independent and Gaussian.
The surface density is in units of objects per $a^2$, where $a$ is the width of the point spread function (see Appendix).
The noise peak distribution is the normalized version of this function.
The surface density has a mode at $\fest \approx 1.33\sigma$ and its shape is well approximated with a Gaussian with standard deviation $\approx 0.835$ for all positive values of $\fest$. 
The shaded (magenta) area highlights the excess density over the Gaussian at negative $\fest$ values.

\begin{figure}
\epsscale{1.2}
\plotone{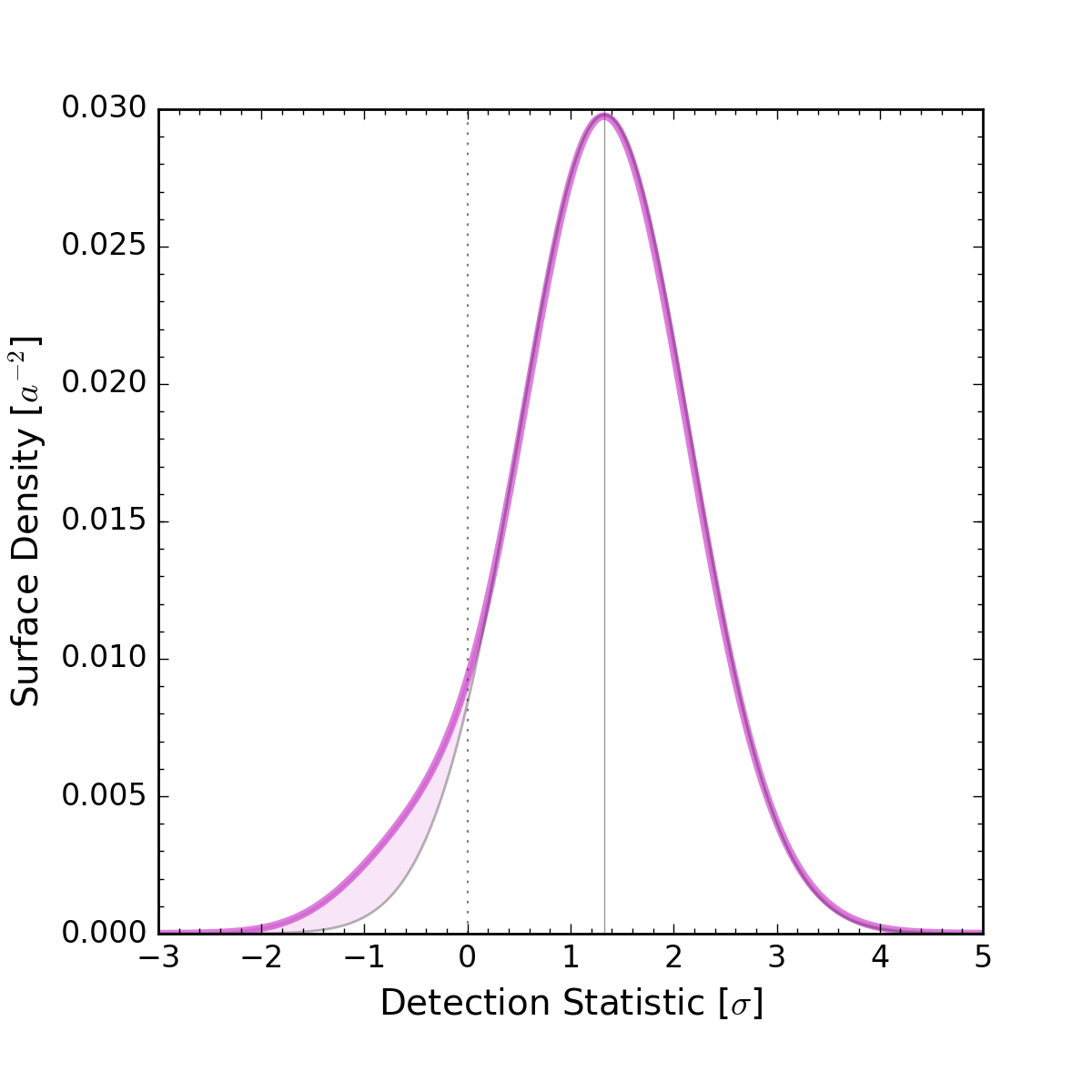}
\caption{The thick (magenta) curve illustrates the surface density of noise peaks as a function of flux in the scenario when the sky noise is white, i.e., has a flat spectrum. The mode of the distribution is at around 1.33$\sigma$ (vertical line) and its shape is well approximated with a Gaussian (thin gray line) for all positive fluxes. The excess density over the Gaussian at negative flux values is shaded (magenta) for clarity.}
\label{fig:surface}
\end{figure}

\begin{figure*}
\epsscale{1.2}
\plotone{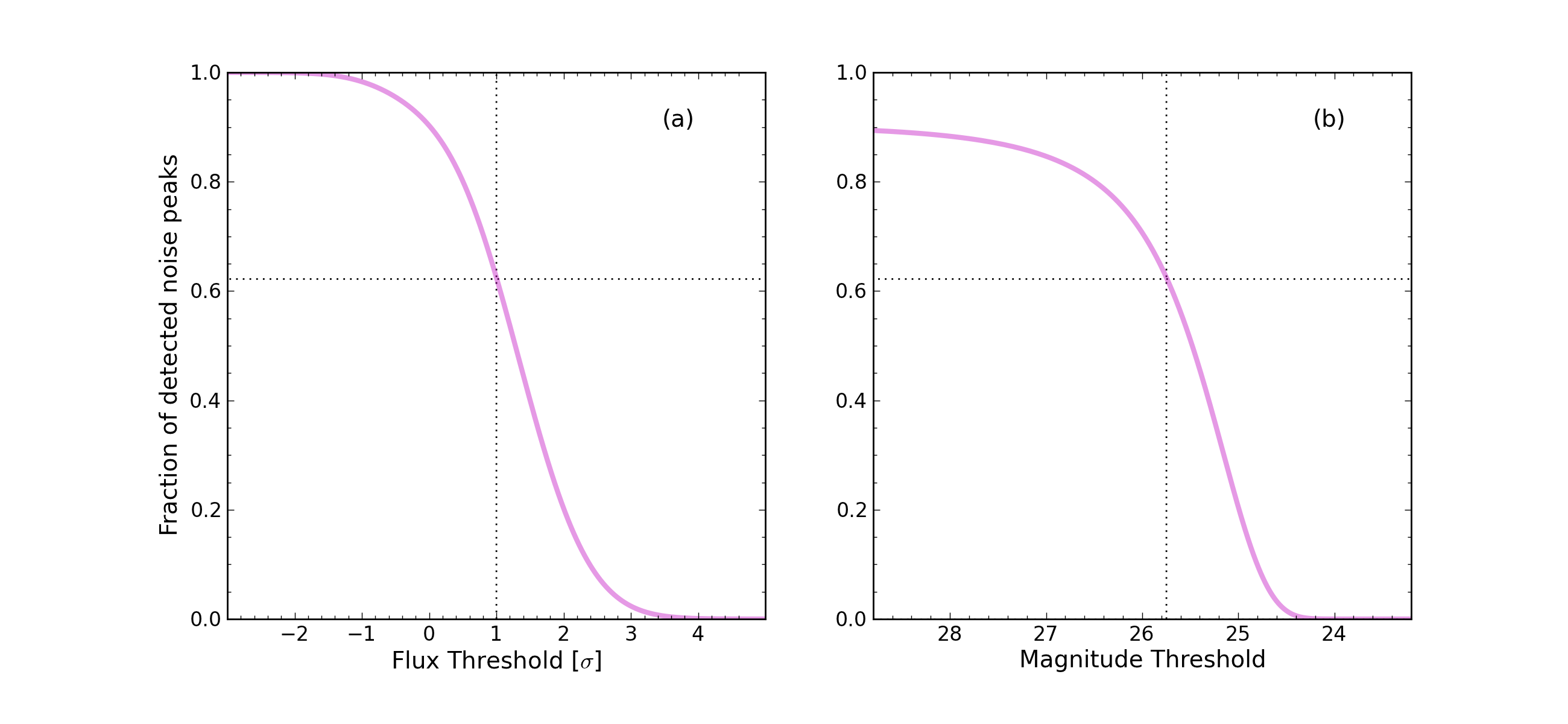}
\caption{The fraction of detected noise peaks drops quickly by raising the threshold. At 1$\sigma$ the value is about 62\% but at 3$\sigma$ it is only a few percent and at 5$\sigma$, which is 24 magnitudes in this case, the detected fraction is negligible. Naively this makes it highly desirable to put a harder constraint on the detection limit but then the opportunity is lost to track fainter sources. The right solution is not this shortcut.}
\label{fig:frac}
\end{figure*}

As noted above, we obtain the probability for detecting a noise peak, $P_N$, by integrating $\npd(\fest)$ above the flux threshold.
Figure~\ref{fig:frac}a shows the results as a function of the flux threshold in $\sigma$ units (left), and on a scale corresponding to an LSST-like magnitude (right; see \S~\ref{sec:disc} for a description of the magnitude scale).
We see that the fraction of noise peaks above threshold is about 62\% at 1$\sigma$, dropping quickly to about a few percent at 3$\sigma$, and becoming negligible at 5$\sigma$.
Based on just this figure, it is tempting to set a high detection threshold to reject such ``ghost peaks'' and keep the catalog of detections nearly pure; but that would mean we lose the opportunity to recover the numerous really faint sources. 

Our multi-epoch approach suggests a different strategy: instead of seeking to make the catalogs for \emph{each} epoch pure, we can adopt a lower single-epoch threshold, relying on the fusion of data across epochs to weed out ghosts.
The marginal likelihood and Bayes factor computations accomplish this data fusion.

The marginal likelihood for the noise hypothesis is a product of the terms for the detections and non-detections:
\begin{equation}
\mlike_{\rm noise} = \left(1\!-\!P_N\right)^{k-n}\,\prod_{\eind \in \dtxn} \npd(\fest_\eind).
\end{equation}
We now have all the ingredients for computing the Bayes factor of Eq.~\ref{eq:Bfac}, providing an objective measure of how much the data prefer a real-source origin to a noise peak origin.

\subsection{Displacements: Including the Astrometry}
\label{sec:astrom}
\noindent
So far we have only used the flux information in the data. 
Genuine sources should have both consistent fluxes and consistent directions across all epochs.
In practice, due to the noise and astrometric errors, the detections of the same object will shift in each exposure, thus the resulting catalogs have to be cross-matched. 
Using a probabilistic method can be to our direct benefit here, enabling straightforward combination of the flux and direction information.

The detections from a real source are all connected, they are just displaced by a random astrometric error; but noise peaks (ghosts) will be independent of each other and their associations can only be by chance.
As we are working under the approximation that the flux and sky position estimates are independent (see Eq.~\ref{eq:eplike}), the Bayes factor using both the photometric and astrometric information factors,
\begin{equation}
B_{\rm{}flux,pos} = B_{\rm{}flux}\cdot{}B_{\rm{}pos}.
\end{equation}
The astrometric cross-match Bayes factor, $B_{\rm{}pos}$, has been derived in \cite{BL15-ARSA} (see Eq.~(17) there, and Eq.~(19) for the tangent plane Gaussian limit that holds for high-precision astrometry).
That work also discusses generalizations that account for proper motion and other complications.

In the following section we assess the discriminative power of multi-epoch source detection by applying it to simulated galaxies and noise peaks, both omitting and including the astrometric data.

\section{Simulations}
\label{sec:disc}

\noindent
We here describe simulations that demonstrate the detection capability of our multi-epoch approach in a setting with known ground truth.
The simulation parameters were chosen to produce data similar to that provided by modern large-scale optical surveys.

\subsection{Simulated Galaxies}
\noindent
We assume that galaxies are brighter than 28 magnitudes and that the 5$\sigma$ detection limit is 24 magnitudes, corresponding roughly to parameters of LSST photometry.
Panel~(b) of Figure~\ref{fig:frac} shows the noise peak detection probability as a function of magnitude based on these parameters, in contrast to the dimensionless presentation in panel~(a).
To compute the marginal likelihood for the source-present hypothesis, we must specify a prior for the source flux, $\flux$.
Here we use a standard faint-galaxy number counts law,
with the number counts following the empirical formula of 
\mbox{$dN\!\propto\!10^{0.4m}dm$}; see \cite{MT00-NumCounts}. 
That approximately translates to the properly normalized population distribution of
\begin{equation} 
\label{eq:ref3}
\pi(f) = \left\{\begin{array}{l l}
           f_L/f^2  & \quad \mbox{if\ $f > f_L$},\\
           0 & \quad \mbox{otherwise},
           \\ \end{array} \right.
\end{equation}
where $f_L$ is the limiting flux that corresponds to the previously defined magnitude limit.

We generate sets of random detections for 20,000 galaxies with true fluxes between 28 and 23 magnitudes by simply drawing $\fest_\eind$ values from a Gaussian centered on the actual fluxes.
We also generate 2,000 ghost detection $\fest_\eind$ values from $\npd(\fest)$ by inverting its cumulative distribution (computed numerically on a grid).
The number of exposures is set to the previously used $k=9$ with a single-epoch flux threshold of just 1$\sigma$, deep in the noise.
In observations with our specified parameters, the number of ghost detections will greatly outnumber the galaxy detections with this low threshold.
The numbers of galaxies and ghosts were chosen to enable display of the distributions of Bayes factors for the two classes of detections (noise and true).

\begin{figure}
\epsscale{1.2}
\plotone{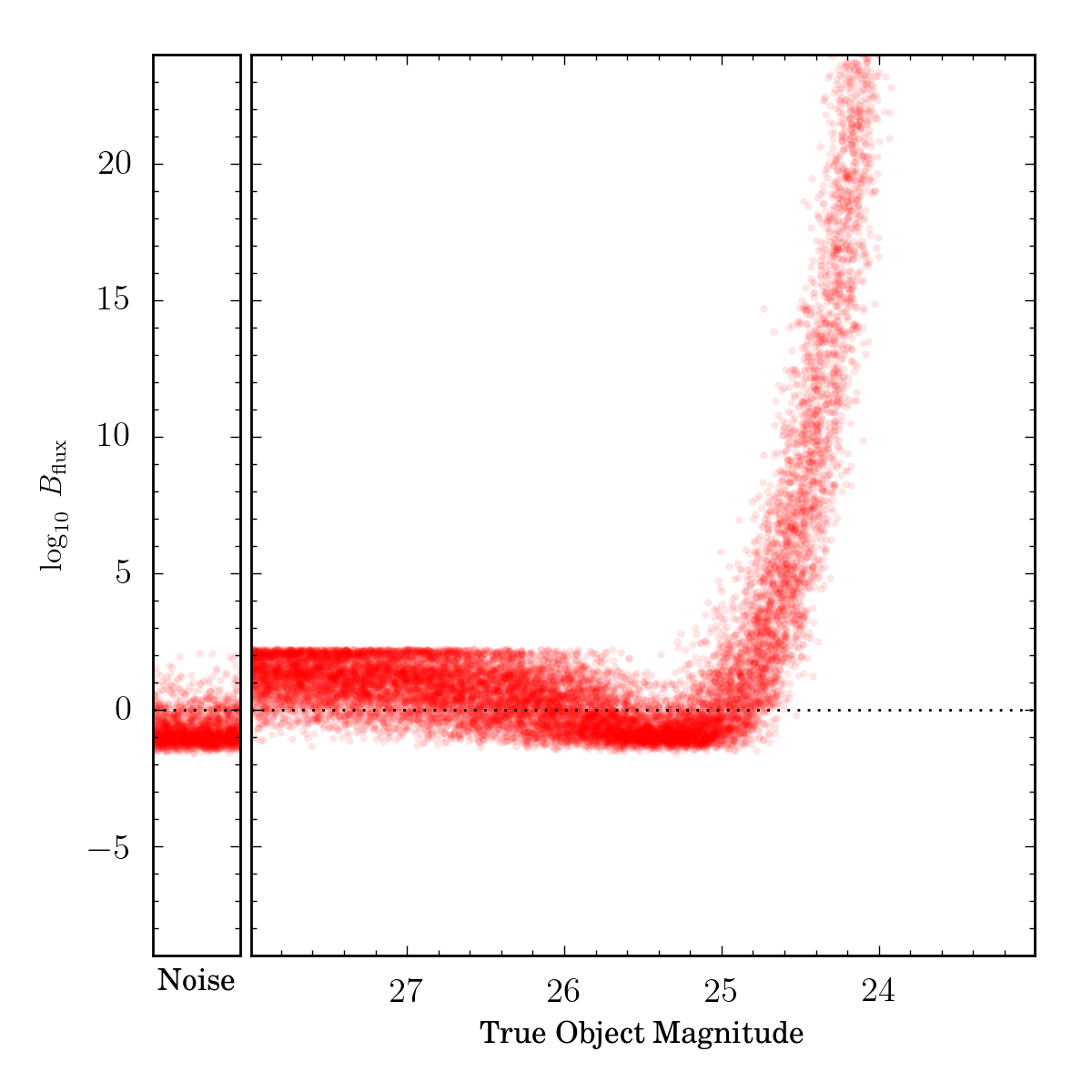}
\caption{Bayes factor distributions for simulated galaxies as a function of their true brightness (right axes), and the same for generated random noise peaks (left axes), using simulated photometric data (ignoring astrometric information).
Points (red) indicate the weight of evidence, quantified via (logarithmic) Bayes factors comparing real-source and noise origins for the simulated photometric data.
The Bayes factor is high for the bright galaxies but is close to unity (logarithm close to zero) for the faint ones.
The Bayes factor distribution for the noise peaks (left axes) peaks below the distribution for the faintest real sources (left region of right axes) because the measured fluxes in noise peaks tend to be in a range that mimics sources between 26 and 25 magnitudes, rather than the dimmest sources.}
\label{fig:bf-photo}
\end{figure}

We first analyze the simulated data considering only the photometric information (i.e., ignoring the directional Bayes factors).
In Figure~\ref{fig:bf-photo} the (red) points represent the resulting Bayes factors for the real sources (right of the double dashed vertical lines) and the noise peaks (on the left).
Superficially, the Bayes factors may appear surprisingly large; even for dim sources the Bayes factors are often $\sim 10^2$, often considered strong evidence in settings where the competing hypotheses are assigned prior odds of unity.
But here, the prior odds for a genuine association vs.\ a noise peak match are extremely small, because chance associations are likely due to the high spatial density of galaxies.
\cite{pxid}, \cite{B13-SCMAXMatch}, and \cite{L13-HierXMatch} discuss how to compute the prior odds in various settings.

Figure~\ref{fig:bf-photo} shows that, as one would expect, the weight of evidence is strong for the bright galaxies but weakens for the faint galaxies.
The smallest Bayes factors arise for galaxies with true magnitudes near 26.5, which corresponds to the mode of the noise peak distribution, $\npd(\fest)$.
Perhaps surprisingly, sources dimmer than magnitude 26.5 can have larger Bayes factors than those with magnitude 26.5.
This happens because $\npd(\fest)$ peaks away from $\fest=0$, i.e., we do not expect noise peaks to have arbitrarily small measured fluxes; the peak-finding process biases the noise peak distribution away from zero flux.
For the weakest detectable sources, the most likely number of detections among the $k=9$ epochs is one.
The flat top of the distribution at the faint end corresponds to very dim sources detected only once, very near threshold.
The smaller Bayes factors in that region of the plot correspond to unlikely larger numbers of detections near the threshold; the discreteness in the number of detections produces a subtle banding in the distribution.

We now consider the astrometric data, by itself.
For simplicity, we assume a constant direction uncertainty of $0.1\arcsec$ for all detections.%
\footnote{In reality, the direction uncertainty will be roughly constant for bright sources, when estimation error is dominated by systematic uncertainty, but for dim source candidates it will grow with magnitude (i.e., as estimated flux decreases).
This will decrease the values of Bayes factors for dim sources, but does not qualitatively impact the findings described below.\label{fn:const-sig}}
We simulate the coordinates for the mock galaxies as follows.
Around the true direction of each object, we randomly generate points from a 2D Gaussian.
This flat sky approximation is excellent in this regime; for such tight scatters, the approximation error is below the limit of the numerical representation of double precision floating point numbers.

The coordinates of noise peaks are generated homogeneously.
The surface density of the ghosts is analytically calculated and its integral above the 1$\sigma$ detection threshold yields \mbox{$\nu\!=\!0.04/\square\arcsec$}.
A simple algorithm is to pick a large enough square, with area $\Omega$, and randomly draw the number of peaks from a Poisson distribution with expectation value $\lambda=\nu\Omega$.
Out of these ghosts, we pick a number equal to the number of flux detections, with locations such that are closest to the center, where the simulated object is placed.

\begin{figure}[t!]
\epsscale{1.2}
\plotone{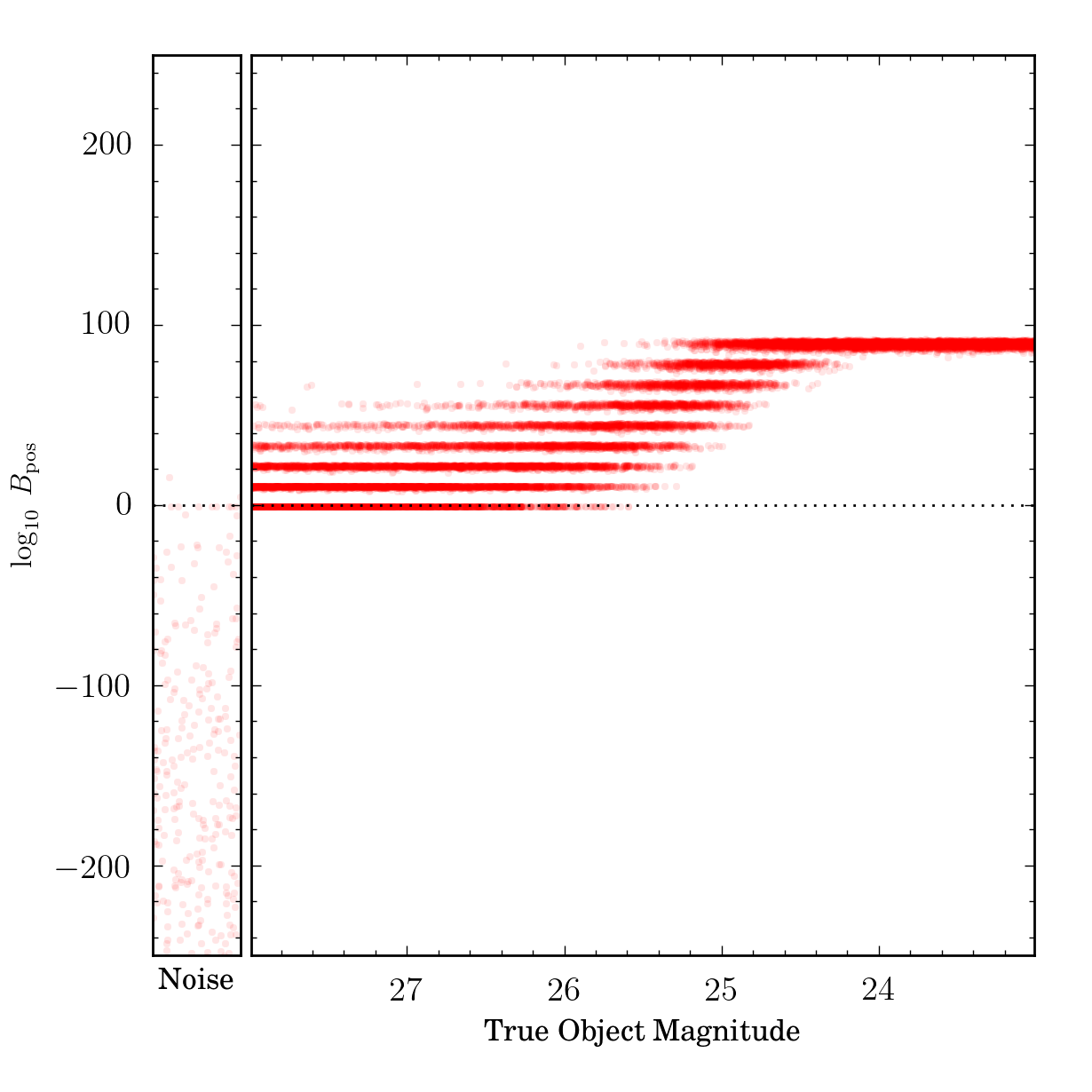}
\caption{Bayes factor distributions as in Figure~\ref{fig:bf-photo}, but here accounting for only astrometric information, i.e., cross-identification across epochs based on the celestial coordinates estimates and uncertainties.
Note the larger range for the Bayes factor axis.}
\label{fig:bf-astro}
\end{figure}

Figure~\ref{fig:bf-astro} shows the distribution of astrometric Bayes factors, for the real (mock) and noise sources.
Note the larger span of the (log) Bayes factor axis.
Banding due to discreteness in the number of detections is now clearly apparent among the true-object Bayes factors; the 9 levels correspond to the different number of detections with the lowest being 1.
Comparing to Figure~\ref{fig:bf-photo}, we see that directional cross-matching is a stronger discriminant between real and noise sources in the dim source regime.
The photometric data grow in importance as sources grow brighter.

The astrometric Bayes factors are essentially constant vs.\ magnitude for a given number of detections among the 9 epochs.
This is a consequence of our simplifying assumption of a constant direction uncertainty.
As noted in footnote~\ref{fn:const-sig}, in real surveys the astrometric precision decreases with increasing magnitude (decreasing flux) in the weak-source regime; this would lead to some decrease in the Bayes factors with increasing magnitude.

\begin{figure}[t!]
\epsscale{1.2}
\plotone{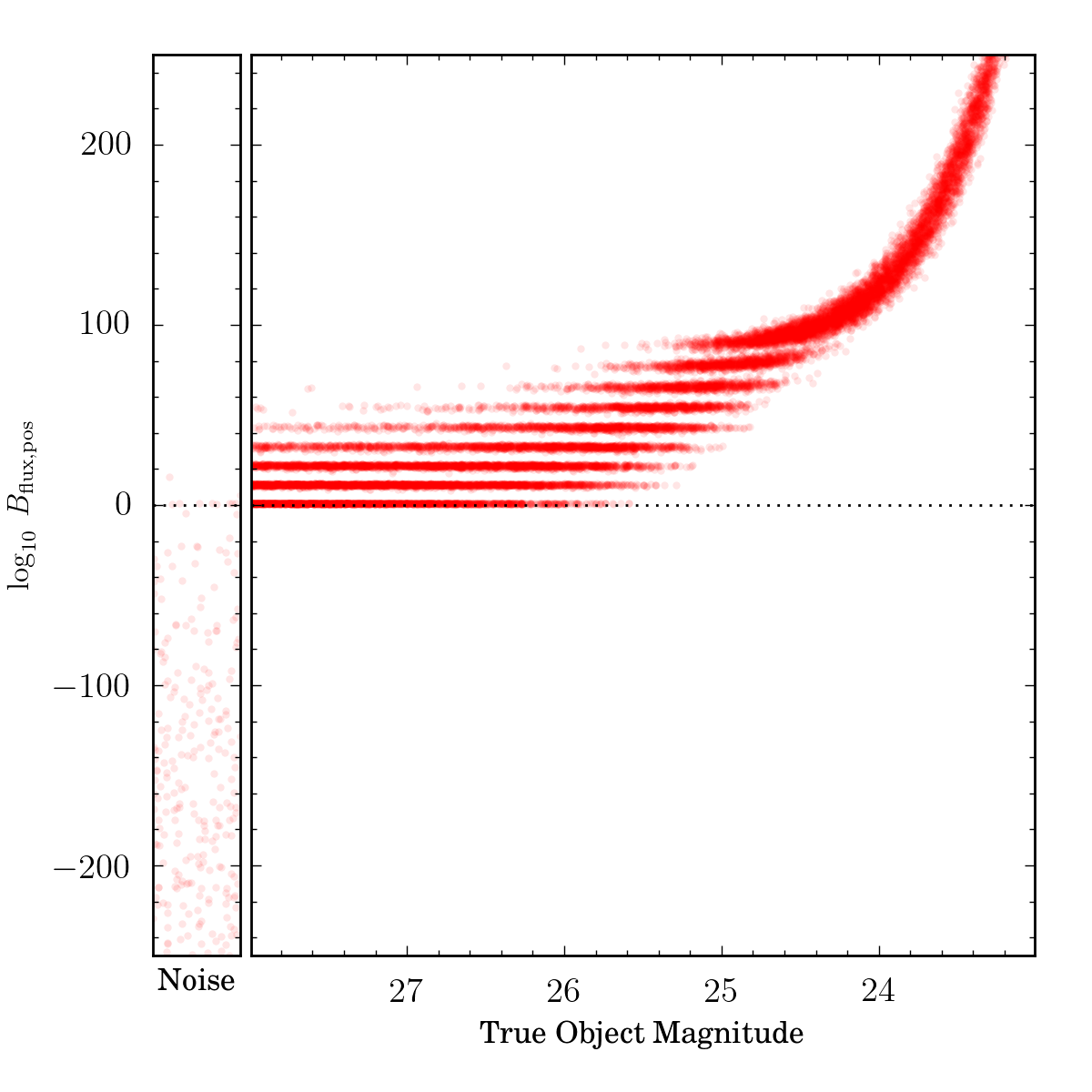}
\caption{Bayes factor distributions as in Figure~\ref{fig:bf-photo}, but accounting for both photometric and astrometric information.
Points (red) indicate the logarithm of the product of the photometric and astrometric Bayes factors.
The astrometric factor dominates for weak sources, and the photometric factor dominates for bright ones.}
\label{fig:bf-both}
\end{figure}

Figure~\ref{fig:bf-both} shows the distribution of Bayes factors, for the real and noise sources, now combining the photometric and astrometric factors.
For the lowest band, corresponding to a single detection, $B_{\rm{}pos}$ is unity by definition (no constraint coming from a single detection), producing the same Bayes factor distribution as in the flux-only calculation.
For multiple detections, the Bayes factors for the true sources are greatly enhanced by including astrometric information (note that the ordinate is logarithmic); just two detections produces quite strong evidence for the true-object hypothesis.
The Bayes factors for the noise peaks have moved to much lower values, due to the low likelihood of directional coincidences.

This computation demonstrates that flux and astrometric catalog data, combined across epochs, can strongly distinguish real objects from spurious detections.
We have not addressed what threshold Bayes factor to use for producing a multi-epoch catalog, or, in Bayesian terms, how to convert Bayes factors into posterior probabilities for candidate objects.
As noted above, when the object population density (on the sky and in flux) is known a priori, the calculation is straightforward (e.g., the prior odds will be proportional to the ratio of true object and noise peak sky densities).
When these quantities are unknown, a possibly complicated hierarchical Bayesian calculation can jointly estimate them and the object properties.
When many objects are detected, an approximate approach, plugging in empirical estimates of the densities based on the data, is likely to suffice, as described in \cite{B13-SCMAXMatch,L13-HierXMatch}.

\color{red}

\color{black}

\section{Summary}
\label{sec:sum}

\noindent
This paper presents an exploratory study of a new, incremental approach to the analysis of multi-epoch survey data, based on fusion of single-epoch catalogs produced using a source detection algorithm with a modest or low threshold.
Although the single-epoch catalogs will include many noise sources (they may even be dominated by them), we show that probabilistic fusion of the single-epoch data can produce interim or final multi-epoch catalogs with properties similar to those expected from catalogs based on image stacking.
The approach is essentially a generalization of cross-matching, where object detection corresponds to identifying a set of candidate sources that match in both flux and direction across epochs.
Using a probabilistic approach directly provides the required quantities, enabling fusion of information both across epochs, and between flux and direction, by straightforward multiplication of the relevant probabilities.
The final quantification of strength of evidence is via marginal likelihoods and Bayes factors; these can be used for final thresholding, or for producing posterior probabilities for source detections when population properties such as sky densities are known or can be accurately estimated (perhaps as part of the catalog analysis).

The Bayes factor compares predictions of the observed data based on true-object and noise-peak hypotheses for the data, and thus requires knowledge of the distribution of noise peaks.
We derive the spatial properties of noise peaks that commonly appear in catalogs.
The flux-dependent surface density of ghosts is asymmetric in flux, skewed toward positive flux values.
It can be accurately approximated by a shifted Gaussian for most practical purposes.

Based on the Bayes factor, sources with single-epoch measured fluxes over a \mbox{3$\sigma$} threshold \mbox{($m\!\simeq\!24.55$ for an LSST-like survey)} start to separate out from the noise peaks when data are combined across just a few epochs.
The evidence for a source becomes very strong once the single-epoch fluxes exceed \mbox{5$\sigma$} (24 mag).
When considering only the flux measurements, the faintest sources are hard to distinguish from the noise peaks with measurements at just a few epochs; but astrometric data (celestial coordinate estimates) greatly help to separate genuine and spurious detections.

We have treated only the case of detection of constant-flux sources.
Detecting variable and transient sources can be accommodated by introducing one or more time series models into the flux matching part of the algorithm.
Models that accurately describe particular classes of sources will produce optimal catalogs, but flexible models---perhaps simple stochastic processes, or even histogram or other partion-based models, with appropriate priors on variability---may suffice for producing general-purpose multi-epoch catalogs for studying variable sources.
This is a potentially complicated generalization of our framework that we plan to explore in future work.
Of course, variable source detection using image stacks is also an open research problem; our framework provides an alternative avenue to address it.

Our exploratory study made simplifying assumptions.
A strong assumption was that source candidates are isolated enough that the image space can be partitioned into patches that have at most one candidate source.
When source candidates are close to each other, the matching across epochs must account for multiple possible source association hypotheses.
Similar complications appear when considering classes of objects that may be comprised of multiple sources per epoch, e.g., radio galaxies.
In other work, we have developed techniques for directional cross-matching in contexts with multiple candidate associations, and with complex object structure and object motion \citep{BL15-ARSA}.
These methods can be extended to include flux matching criteria to generalize the multi-epoch detection framework described here.

The strategy we have described is quite different from conventional approaches to producing survey catalogs.
Implementing it will raise new processing and database management challenges; users of the resulting catalogs will need to think about catalogs in a different way.
In particular, a low-threshold single-epoch catalog will contain many spurious sources; with a low enough threshold, the spurious sources will greatly outnumber real sources.
However, evidence mounts quickly as catalogs are merged.
If interim catalogs are produced consecutively, cumulative culling of early single-epoch catalogs could reduce the storage burden for catalogs subsequent to the first catalog.
Such issues, and the generalizations described above, will be topics for future study.

\acknowledgements{}
The authors gratefully acknowledge valuable and inspiring discussions with Andy Connolly and Robert Lupton on various aspects of the topic.
This study was supported by the 
NSF via grants AST-1412566 and AST-1312903, and the NASA via the awards NNG16PJ23C and STScI-49721 under NAS5-26555.
%

\appendix

\section{Peaks of Two-Dimensional Random Fields}
\noindent
Consider a two dimensional Gauusian random field $f(\rr)$, with a known power 
spectrum. Its gradient would be $\hh$, and the second derivative tensor $g$. 
We would like to find out the density of peaks of this field above a certain 
height. We will follow the procedure outlined in \citet{bbks}.

We will expand the field and its gradient to second order around a peak at the 
position $\rr_p$:
\begin{eqnarray}
	f(\rr) = f(\rr_p)+\frac{1}{2} g_{ij}(\rr_p) (\rr-\rr_p)_i  (\rr-\rr_p)_j\\
	h_i(\rr) = (\rr-\rr_p)_j g_{ij}(\rr_p),
\end{eqnarray}
where we already use the fact that the gradient of the field at a peak is zero, 
i.e. $h_i(\rr_p)=0$. Provided that $g$ is non-singular at $\rr_p$, we can express 
$\rr-\rr_p$ from the second equation:
\begin{equation}
	\rr-\rr_p = g^{-1}(\rr_p) \hh(\rr_p).
\end{equation}
We can write a Dirac delta that picks all extremal points of $f$ as
\begin{equation}
	\delta^{(2)}(\rr-\rr_p) = |\det g(\rr)| \delta^{(2)}[\hh(\rr)].
\end{equation}
This expression turns a continous random field, defined at all points over our 
two-dimensional space into a discrete point process, that of the extremal points 
of the field,
\begin{equation}
	n_{\rm{}ext}(\rr) = |\det g(\rr)| \delta^{(2)}[\hh(\rr)].
\end{equation}

In order to pick the peaks of the Gaussian random field we will also need to have 
a negative definite $g$. If we only want peaks of a certain height, we need to 
calculate the appropriate ensemble average of this density over the constrained 
range of the variables.

We have six random variables, the field $f$, the three components of the symmetric 
$g$ tensor, and the two components of the gradient $\hh$. The correlations can be 
computed in a straight-forward manner, given the power spectrum of the field. The
gradient is uncorrelated with both the field and the second derivatives, due to the
parity of the Fourier representation. Let us denote the correlation matrix of the
field and the Hessian by $C$, and that of the gradient as $H$. Furthermore, let us 
define the different $k$-moments of the power spectrum characterizing the field as
\begin{equation}
	\sigma_n^2 = \frac{1}{(2\pi)^2} \int d^2k\, k^{2n} P(k).
\end{equation}

We can now explicitely write down the correlation matrix $C$ of 
${\bf v} = (f,g_{11},g_{12},g_{22})$ and $H$ for $\hh= (h_1, h2)$, as
\begin{equation}
	 C = \left(
		\begin{array}{cccc}
			\sigma_0^2 & -\sigma_1^2/2 & 0 & -\sigma_1^2/2\\
			-\sigma_1^2/2 & 3\sigma_2^2/8 & 0 & \sigma_2^2/8\\
			  0 & 0 & \sigma_2^2/8  & 0\\
			-\sigma_1^2/2 & \sigma_2^2/8 & 0 & 3\sigma_2^2/8
		 \end{array}\right),
\end{equation}
\begin{equation}
	 H = \left(
		\begin{array}{cccc}
			-\sigma_1^2/2 & 0 \\
			0 & -\sigma_1^2/2
		 \end{array}\right).
\end{equation}
With these we can write the multivariate Gaussian distribution using the inverse 
of the correlation matrix as a product of two independent distributions
\begin{equation}
	dP =\exp\left(-\frac{\vv^T C^{-1}\vv}{2}\right)
  			\frac{d^4 \vv} {(2\pi)^2|C|^{1/2}}
		 \exp\left(-\frac{\hh^T H^{-1} \hh}{2}\right) 
			\frac{d^2 \hh} {2\pi|H|^{1/2}}.
\end{equation}
Before we proceed further, the second derivative tensor can be described more conveniently
with the two eigenvalues $\lambda_1,\lambda_2$ and a rotation angle,
$\phi$, as follows:
\begin{eqnarray}
	g_{11} =& \lambda_1 \cos^2\phi +\lambda_2 \sin^2\phi,\\
	g_{12} =& (\lambda_1-\lambda_2) \sin\phi \cos\phi,\\
	g_{22} =& \lambda_1 \sin^2\phi + \lambda_2 \cos^2\phi.
\end{eqnarray}
For simplicity let us introduce the dimensionless variables 
$x=(\lambda_1+\lambda_2)/\sigma_2$, the trace of the second derivative tensor, 
$y=(\lambda_1-\lambda_2)/\sigma_2$, and $z = f/\sigma_0$. The Jacobian of the 
transformation from $(f,g_{11},g_{12},g_{22})$ to $(z,x,y,\phi)$ is
\begin{equation}
		J = \sigma_0\sigma_2^3 y/2.  
\end{equation} 
Let us also introduce the dimensionless $\gamma$ and the characteristic scale 
$R$ as
\begin{equation}
	\gamma = \frac{\sigma_1^2}{\sigma_0\sigma_2},
		\qquad R^2 = \frac{\sigma_1^2}{\sigma_2^2}.
\end{equation}
The quadratic form containing $\vv$ in the exponent can be written with the new 
variables as
\begin{equation}
	Q = v^T C^{-1} v = \left(
		2 y^2	+\frac{x^2 + 2\gamma x z + z^2}{1-\gamma^2}\right).
\end{equation}
The determinants of $C$ and $H$ are 
\begin{equation}
	|C| = \frac{1}{64}(1-\gamma^2)\sigma_0^2\sigma_2^6,\qquad
	|H| = \frac{1}{4} \sigma_1^2.
\end{equation}
In these variables, the unconstrained probability distribution for 
$(x,y,z,\phi)$ becomes
\begin{equation}
	dP = \frac{4y}{(2\pi)^2 \sqrt{1-\gamma^2}} \exp(-\frac{1}{2}Q) 
		\,dx\ dy\ dz\ d\phi.
\end{equation}
In order to properly handle the symmetries of the problem, we can assume that 
$\lambda_1\geq\lambda_2$. Then still any $(\lambda_1,\lambda_2)$ pair can be 
mapped onto itself by a 180 degree rotation, so the valid range of $\phi$ is 
$(0,\pi)$.  Since none of the terms depend on $\phi$, we can integrate 
over $\phi$, resulting in
\begin{equation}
	dP =  
		 \exp\left[-\frac{x^2 + 2\gamma x z + z^2}{2(1-\gamma^2)}\right] 
			\frac{\,dx\ dz}{2\pi \sqrt{1-\gamma^2}}
		\left( e^{-y^2}\,2y\ dy\right).
\end{equation}
The constraint $\lambda_1>\lambda_2$ maps onto $0<y$. If we perform the 
integration over $y>0$, and $x,z$ over $(-\infty,\infty)$, we get 1, as we 
should, for the unconstrained probability for a general point.

As we introduce the peak constraints, we need to first consider the impact
on the gradient. The constrained probability distribution
is
\begin{equation}
	dw_x = \exp\left(-\frac{\hh^T H^{-1} \hh}{2}\right) 
			|\det g| \delta^{(2)}[\hh]			
			\frac{d^2 \hh} {2\pi|H|^{1/2}}.
\end{equation}
After integrating over $d^2\hh$ we get the extremum weight
\begin{equation}
	w_x = \frac{|\det g|}{2\pi|H|^{1/2}} =\frac{x^2-y^2}{4 \pi R^2}.
\end{equation}
This will multiply the unconstrained probability for the density of extremal 
points of the random field, 
\begin{equation}
		dn_{\rm{}pk}=\left[\frac{(x^2-y^2)y}{2\pi R^2} e^{-y^2}\,dy\right]
		\exp\left[-\frac{x^2 + 2\gamma x z + z^2}{2(1-\gamma^2)}\right] 
			\frac{\,dx\ dz}{2\pi \sqrt{1-\gamma^2}}.
\end{equation}
For a peak both eigenvalues of the second derivative tensor must be negative. 
In the rotated coordinates $(x,y)$, this means that $x<0$, and $0<y<-x$.
We can easily integrate over the allowed range of $y$ next, yielding
\begin{equation}
	\int_0^{-x} dy\, y\/ (x^2-y^2)\, e^{-y^2} 
		=\frac{1}{2}(x^2-1+e^{-x^2}).
\end{equation}
We are left with
\begin{equation}
	dn_{\rm{}pk} = \frac{(-1 + x^2 +e^{-x^2})}{4\pi R^2}  
	\exp\left[-\frac{x^2 + 2 x z \gamma + z^2 }{2 (1-\gamma^2)}\right]
	\frac{\,dx\ dz}{2\pi \sqrt{1-\gamma^2}}.
\end{equation}
Let us introduce the function $B(x)$ as
\begin{equation}
	B(s,b) = \sqrt{\frac{\pi}{b}} \exp\left({\frac{s^2}{2b}}\right)
	\left[1+ {\rm erf}\left( \frac{s}{\sqrt{2b}}\right)\right].
\end{equation}
Evaluating the integral over $-\infty<x\leq 0$ in Mathematica, we obtain
\begin{equation}
 n_{\rm{}pk}(z)= 
		\frac{e^{-\frac{s^2}{2\gamma^2}}}{8\pi^2R^2}
		\Bigg[
			(1-\gamma^2)s +[s^2 - \gamma^2 (1 + s^2)] B(s,1)
			+B(s, 3- 2\gamma^2),
		\Bigg],
\label{eq:npk}
\end{equation}
with
\begin{equation}
	s = \frac{\gamma z}{\sqrt{1-\gamma^2}}.
\end{equation}
We get the full surface density of noise peaks, $\lambda$, by integrating the conditional surface density in Eq.~\ref{eq:npk} over all peak heights $z$:
\begin{equation}
	\lambda = \int_{-\infty}^{\infty} n_{pk}(z) dz.
\end{equation}
Finally, we need to evaluate the shape parameter $\gamma$. Assume that the window 
function applied to the random field is a Gaussian with a scale $a$,
\begin{equation}
	w(r) = \frac{1}{2\pi a^2} \exp\left(-\frac{r^2}{2 a^2}\right).
\end{equation}
Its Fourier transform is also a Gaussian,
\begin{equation}
	W(k) = \exp\left(-\frac{k^2a^2}{2}\right).
\end{equation}
We model the sky noise as a white noise with a flat spectrum. Thus the
correlations in the measured random field are determined by the window function,
i.e.,
\begin{equation}
	P(k) = A\,|W(k)|^2 = A \exp\left(-k^2a^2\right).
\end{equation}
With this power spectrum it is straightforward to compute the scale and the shape parameters as
\begin{equation}
	\gamma^2 = \frac{1}{2}, \qquad R^2 = \frac{a^2}{2}.
\end{equation}
With this choice of PSF and $\gamma$, we get $s=z$ in Eq.~\ref{eq:npk}.

\begin{figure}
\epsscale{0.95}
\plottwo{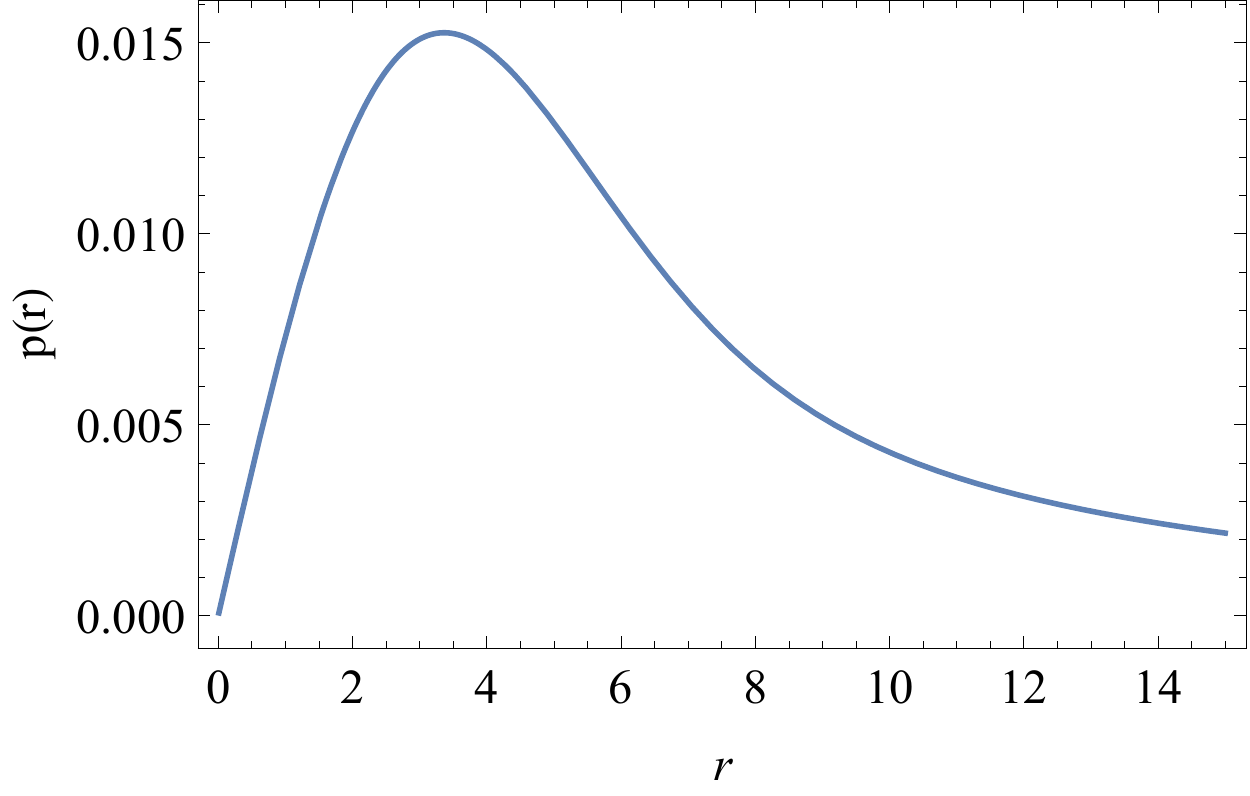}{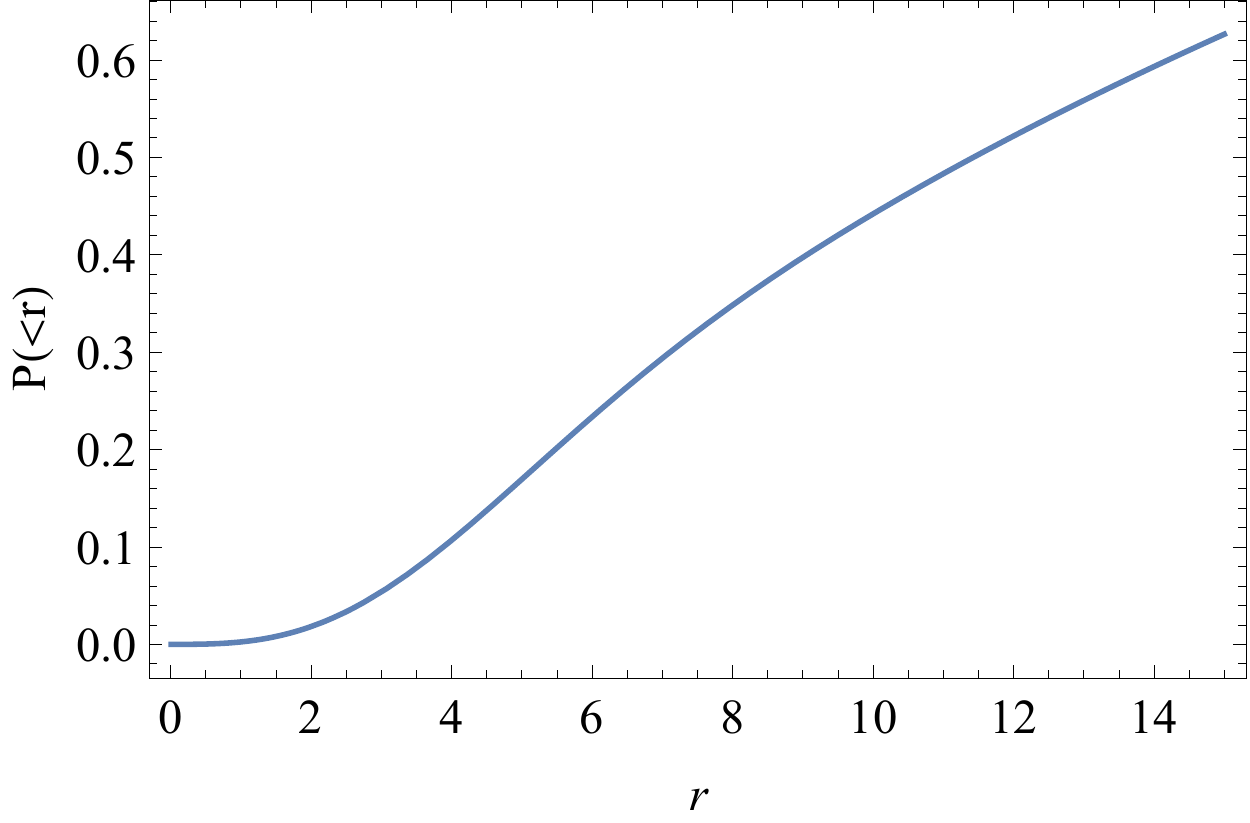}
\caption{The differential and cumulative probabilities of having a noise peak at or within a radius $r$, integrated over all heights. The radius is shown in units of $a$, the PSF scale.
}
\label{fig:pk-shift}
\end{figure}

Now we are in a position to compute the probability that a noise peak is within a radius $r$ of our point of interest located at the origin. The spatial distribution of the noise peaks is described by a Poisson process with the surface density $\lambda$. 
The cumulative probability that the peak is within a radius $r$ is given by the well-known expression
\begin{equation}
	P_{pk}(<r|z) = 1-\exp[-\lambda(z) r^2 \pi].
\end{equation}
The differential probability is given by its derivative with respect to to $r$, as 
\begin{equation}
 p_{pk}(r|z) = 2\pi r \lambda(z) \exp[-\lambda(z) r^2 \pi]
\end{equation}
Both of these probabilities are shown on Fig.~\ref{fig:pk-shift}. The differential probability starts off around the origin scaling with $r$, due to the available area (phase space for configuration).
This in turn causes the cumulative function to rise as $r^2$, resulting in a very small probability ($<0.03$) that a noise peak will appear within a PSF scale.
Thus we can safely ignore noise peaks as a major contributor to false detections at a significant level.

\end{document}